\documentclass{ifacconf}

\usepackage{natbib}        
\usepackage{graphicx} 
\usepackage{ mathtools, amssymb, mathrsfs,mathabx}  
\usepackage{enumerate} 
\usepackage{array,multirow}
\usepackage[table]{xcolor}
\usepackage{ diagbox}

\begin{document}
\begin{frontmatter}

\title{Data-driven model reduction: An identification perspective using Generalized Bases of Orthonormal Functions} 

\author[Vau]{Bernard Vau} 
\author[Bourles]{Henri Bourlès} 

\address[Vau]{EXAIL, 
   94380 Bonneuil-sur-Marne, France (e-mail: bernard.vau@exail.com)}
\address[Bourles]{CNAM, SATIE, France (e-mail: henri.bourles@lecnam.net)}

\begin{abstract}                
This paper proposes a data-driven model order reduction approach based on system identification concepts. A revised equation error formulation is introduced, in which the model is parameterized using Generalized Bases of Orthonormal Transfer Functions (GBOFs). The generalized reduction problem is formulated in the Hambo frequency domain associated with the GBOF transform. The frequency-domain fit of the reduced-order model is adjusted through the selection of the basis poles. A fit indicator depending on these poles is introduced, providing valuable heuristic guidance for their selection.
\end{abstract}


\end{frontmatter}

\section{Introduction}

While dynamical models derived from physics are often extremely complex, the need for reduced-order models suitable for simulation is crucial in many engineering fields. In control engineering, an overly complex model is often inappropriate for controller synthesis. The field of model order reduction (MOR) has now reached maturity (\cite{r29}), (\cite{r28}). In most cases, MOR is performed on a known model. However, the current data-driven era has recently generated significant interest in data-driven model order reduction methods. These approaches can be divided into several categories depending on the available system data. The present paper focuses on input-output data-driven approaches, for which a substantial body of literature already exists (\cite{r30}), (\cite{r31}), (\cite{r32}). \\

\noindent Obtaining a model, whether reduced or not, from experimental data is, by definition, a system identification task. For this reason, the system identification community has an important contribution to make to this field. We restrict here our attention to linear discrete-time SISO (single-input single-output) systems. The main challenge associated with the estimation of a reduced-order model from data is to control the quality of the model fit in the frequency domain. In the noise-free case, the model approximation bias depends on the input spectrum and the flexibility of the chosen model structure. Such a model can be obtained directly by minimizing the sum of squared equation errors, which is straightforward to implement due to the convexity of the problem. The minimization of the sum of squared equation errors was also addressed many years ago by the MOR community from a non-data-driven perspective (\cite{r27}), but does not seem to have received widespread attention. As shown in the following, the standard equation error method is of limited interest, since it provides an effective approach to model reduction only when the reduced-order model is expected to provide a satisfactory approximation over a linear frequency scale, which is seldom the case in MOR. \\

\noindent The purpose of this paper is to propose what we call a data-driven generalized equation error method, in which the reduced-order model is parameterized using Generalized Bases of Orthonormal Transfer Functions (GBOFs), introduced in the control community in the 1990s (see (\cite{r1})). GBOFs generalize the usual Fourier basis $z^{-1}, z^{-2}, \ldots$, as well as the Laguerre and Kautz bases, and have their own associated transform: the Hambo transform. The effect of using such a parameterization in what we call a generalized ARX (Auto-Regressive with eXogenous input) model is equivalent, in the frequency domain, to transforming the initial equation error problem into a warped frequency scale, called the Hambo frequency scale, introduced in (\cite{r24}) (see also Chapter 8 of (\cite{r1})). After expressing the generic form of the criterion to be minimized in the Hambo frequency domain, we use the warping rate, called the function $\chi$ in (\cite{r2}), between the usual logarithmic frequency scale and the Hambo frequency scale, to interpret the effect of the basis poles selection on the fit of the reduced-order model. It turns out that this function $\chi$ is a remarkable indicator of this effect. The remainder of this paper is organized as follows:  Section II proposes a short review on GBOFs and their transforms, in Section III the reduction problem is stated, Section IV proposes and analyses an initial equation error reduction. In Section V, the generalized equation error method is detailed with simulation examples. At last the issue of additive stochastic noise on output data is briefly discussed.

\section{Review on GBOF, Hambo transforms, and the Hambo frequency}

Before describing the novel algorithm proposed here, it is necessary to recall some basic notions about GBOFs and the Hambo transform. The generalized bases of orthogonal transfer function are generated from an all-pass transfer function (Blaschke product) $G_{b}(z)=\prod_{j=0}^{n_{p}-1}\frac{-z.\bar{p}_j+1}{z-p_{j}}$ where $p_j$ $(\lvert p_j\rvert<1)$ are the basis poles and $n_p$ the basis poles number. There exists an orthonormal state space realization such that $G_{b}(z)=D_b+C_{b}\left(zI-A_b\right)^{-1}B_b$. The orthonormal transfer function vectors $V_k(z)$ with $k=1,2, \cdots$ and of size $(n_p,1)$ are given by the relation $V_{k}(z)=\left(zI-A_b\right)^{-1}B_{b}G_{b}^{k-1}(z) $. Orthonormality between these functions holds, because of the orthonormal state space realisation of $G_b(z)$. Particular configurations of $n_p$ and $p_k$ correspond to well-known cases: $n_p=1, p_{0}=0$ is the classical $z^{-1}, z^{-2}, \cdots$ basis, and  $n_p=1, \lvert p_0 \rvert <1$ correspond to the Laguerre basis. \newline
\noindent The Hambo operator $\lambda$ is given by $\lambda^{-1}=G_b(z)$. The mapping $\lambda \mapsto z$ is $n_p$ valued in a domain including the unit circle, and the $z_j$ $j=1,\cdots n_p$ so that $\lambda^{-1}=G_b(z_j)$ are the eigenvalues of $\mathscr{N}(1/\lambda)$ with $\mathscr{N}(\lambda)=A_b+(\lambda-D_b)^{-1}C_b$. \\
\noindent A transform is associated to GBOFs: the Hambo transform. More exactly there exists a Hambo signal transform and a Hambo operator transform. \\
\noindent 1) Hambo signal transform: set $v_1(t)=A_b^{t-1}B_b$ and  $v_{k+1}(t)=G_b(q)v_k(t)$, with $k \in \mathbb{N}$. The Hambo signal transform of a sequence $\{x(t)\} \in l_2$ (scalar) is $\{\breve{x}(k) \}$ with $\breve{x}(k)=\sum_{t=\mathbb{N}}v_k(t)x(t)$. Note that for a basis having more than one pole $(n_p>1)$, $\breve{x}(k)$ is a vector of size $(n_p,1)$. \\
\noindent 2) Hambo operator transform: The Hambo operator transform of a rational transfer function $H \in \mathbf{H}_2$ is given by $\tilde{H}(\lambda)=\sum_{\tau=0}^{\infty}h(\tau)N^{\tau}(\lambda)$, where $\{ h(\tau) \}$ is the impulse response of $H$. In case of a multipole basis $(n_p>1)$, $\tilde{H}(\lambda)$ is therefore a square matrix of size $(n_p,n_p)$.\\
\noindent For any transfer functions $H_1, H_2 \in \mathbf{H}_2$, the following properties hold: 

\begin{equation}
\left( \widetilde{H_1H_2} \right)(\lambda)=\tilde{H}_1(\lambda)\tilde{H}_{2}(\lambda)=\tilde{H}_{2}(\lambda)\tilde{H}_{1}(\lambda) 
\label{prop1}
\end{equation}

\noindent and for any $a,b \in \mathbb{R}$, one has
\begin{equation}
 \left ( \widetilde{aH_1+bH_2} \right)(\lambda)=a\tilde{H}_1(\lambda)+b\tilde{H}_1(\lambda) 
 \label{prop2}
 \end{equation}
 
 \noindent Moreover,
 \begin{equation}
 \widetilde{\left(H^{-1}\right)}=\tilde{H}^{-1}
 \label{prop3}
 \end{equation}

\noindent Since $G_b$ is all-pass, a function expressing the phase of $G_b(e^{i\omega})$, called the function $\beta$ was introduced in (\cite{r24}). It can be associated to the operator $\lambda$, such that $\beta: [0,2\pi[\to [0,2\pi n_p[$ with $G_b(e^{i\omega})=e^{-i\beta(\omega)}$ for all $\omega \in  [0,2\pi[$, and $\lambda^{-1}=e^{-i\beta(\omega)}$. The Hambo frequency $\omega_{\lambda}$ is by definition $\omega_{\lambda}=\beta(\omega)$, and the mapping $\mathbb{T} \to \: [0, 2\pi n_p[$ : $\lambda \mapsto \omega_{\lambda}$ is $n_p$-valued ($\mathbb{T}$ being the unit circle). It is shown in (\cite{r24}) and chapter 8 of (\cite{r1}) that

\begin{equation}
\omega_{\lambda}=\sum_{k=0}^{n_p-1}\eta_k+2\arctan\left((\frac{1+\rho_k}{1-\rho_k})\tan(\frac{\omega-\eta_k}{2})\right)
\label{eq_11}
\end{equation}

\noindent  where $p_k=\rho_ke^{i\eta_k}$. \\
\noindent The derivative of function $\beta$ plays a great role: this is the warping rate from $\omega$ to $\omega_{\lambda}$, and it is shown in (\cite{r1}) (Chap.3) that 

\begin{equation}
\beta^{'}(\omega)=\frac{d\omega_{\lambda}}{d\omega}=V_1^{T}(e^{i\omega})V_1(e^{-i\omega})
\label {eq_beta_prime}
\end{equation}

\noindent moreover, $V_1^{T}(e^{i\omega})V_1(e^{-i\omega})$ is a particular expression of the reproducing kernel of the Hilbert space associated with GBOFs. One adds another property: From (\cite{r1}) (section 3.3.10), one has
\begin{equation}
\tilde{H}(\lambda)=\sum_{j=1}^{n_p}\frac{V_1(1/{z_j})V_1^{T}(z_j)}{V_1^{T}(z_j)V_1(1/z_j)}H(z_j)
\label{eq_18}
\end{equation}
\noindent where $z_j$ are the eigenvalues of $\mathscr{N}(1/\lambda)$.
\noindent This can be expressed in the unwrapped frequency scale $\omega_\lambda$ yielding
\begin{equation}
\tilde{H}(\omega_\lambda)=\left. H(e^{i\omega})\frac{V_1(e^{-i\omega}))V_1^{T}(e^{i\omega}))}{V_1^{T}(e^{i\omega}))V_1(e^{-i\omega}))}\right |_{\omega=\beta^{-1}(\omega_\lambda)}
\label{eq_19}
\end{equation}
\noindent which entails
\begin{equation}
 \lVert \tilde{H}(\omega_\lambda) \rVert_2^{2}= \lvert H(e^{i\omega}) \rvert^{2}_{\omega=\beta^{-1}(\omega\lambda)}
 \label{prop4}
 \end{equation} 
 
 \noindent $\tilde{H}(\omega_\lambda)$ is defined in function of $\omega_\lambda$ modulo $2n_p\pi$.

\section{Problem statement}

One seeks to obtain a reduced model from the full order, discrete-time and SISO model  $G_0(q, \theta_0)$, excited by an input sequence $\{u(t)\}$. One assumes that the system is corrupted by a centered output error noise sequence $\{v(t)\}$ resulting from the filtering of a white noise sequence ${e(t})$, filtered by a bicausal filter $W_0(q)$, such that the output sequence $\{y(t)\}$ can be expressed as
\begin{equation}
{y(t)=G_0(q,\theta_0)u(t)+v(t)=G_0(q,\theta_0)u(t)+W_0(q,\rho_0)e(t)} \\
\label{e1}
\end{equation}
 \noindent In \eqref{e1} $q$ is the forward shift operator, $\theta_0$, $\rho_0$ the true system parameters vector. The reduced order model $G_r(q)$ is parameterized with GBOFs such that
 \begin{equation}
G_r(q,\theta)=\frac{\sum_{k=1}^{n_a} \nu_k^TV_k(q)}{1+\sum_{k=1}^{n_a} \mu_k^TV_k(q)}=\frac{B(q)/A_p(q)}{A(q)/A_p(q)} 
\label{eq_G}
\end{equation}
\noindent where the monic polynomial $A_p(q)$, called here the characteristic polynomial, is defined as 
\begin{equation}
A_p(q)=\prod_{k=0}^{n_p-1}(q-p_k)^n
\label{eq_Ap}
\end{equation}

\noindent and where $p_1,p_2, \cdots$ are the basis poles of GBOFs, and $n\in \mathbb{N}$ with $n=na/n_p$.
\noindent The parameter vector is given by
\begin{equation}
\theta^T(t)=[\mu_1^T \; \mu_2^T \; \cdots  \nu_1^T \; \nu_2^T \; \cdots ]
 \label{eq_param_vec_HOLOE}
 \end{equation}

 \noindent One defines an equation error
 \begin{equation}
\varepsilon_{LS}(t+1,\theta)=y(t+1)-\theta^T\phi_{LS}(t,\theta)
\label{eq_LS}
\end{equation}
 
 \noindent and
 \begin{multline}
   \phi^T(t)=\left[ -V_1(q)y(t+1) \; -V_2(q)y(t+1) \cdots \right.  \\
  \left. V_1(q)u(t+1) \; V_2(q)u(t+1) \cdots  \right] 
  \end{multline}

 \noindent A very simple manner to obtain a reduced order model from data consists in minimizing the criterion J ($N$ being the number of available samples) which is convex, because of the linearity in \eqref{eq_LS}
 \begin{equation}
 J(\theta)=\sum_{t=1}^N\varepsilon_{LS}^2(t)
 \label{eq_crit_temp}
 \end{equation}
 
\noindent This is a filtered equation error. Note that minimizing an equation error for model reduction is absolutely not new, since this was considered (in a non-data-driven perspective) in (\cite{r27}).

\section{The initial equation error problem}
In case of a noise-free problem ($v(t)=0)$, and if the basis poles of GBOFs are set to $0$, one verifies easily that

\begin{equation}
\varepsilon_{LS}(t)=A_r(q,\theta)\left(G_0(q)-G_r(q,\theta)\right) u(t)
\label{eq_err_temp}
\end{equation}

\noindent and if $G_r(q,\theta)$ is invertible one can write
\begin{equation}
\varepsilon_{LS}(t)=G_r^{-1}\left(G_0(q)-G_r(q,\theta)\right)B_r(q,\theta)u(t)
\label{eq_err_temp2}
\end{equation}

\noindent and by switching in the frequency domain ($\Psi_u(\omega)$ being the Power Spectrum Density of $\{u(t)\}$)
\begin{multline}
{J(\theta)=\frac{1}{2\pi}\int_{-\pi}^{\pi}}\\
{\left|
G_r^{-1}(e^{i\omega},\theta)
\left(G_0(e^{i\omega})-G_r(e^{i\omega},\theta)\right)\right|^{2}
\left|B_r(e^{i\omega},\theta)\right|^{2}
\Psi_u(\omega)\,d\omega}
\label{eq_crit_with_GBOF2}
\end{multline}

\noindent the term $G_r^{-1}(e^{i\omega},\theta)
\left(G_0(e^{i\omega})-G_r(e^{i\omega},\theta)\right)$ can be considered as a relative error, and assuming that  $G_r(q,\theta)$ is sufficiently "close" to $G_0(q)$ (a situation which occurs if the model has enough flexibility) so that
\[\lVert G_r^{-1}(e^{i\omega},\theta)
\left(G_0(e^{i\omega})-G_r(e^{i\omega},\theta)\right) \rVert \ll 1\] 
\noindent one can write

\[\left|\frac{G_0(e^{i\omega})-G_r(e^{i\omega},\theta)}{G_r(e^{i\omega},\theta)} \right | \approx \left |    log(G_0(e^{i\omega}))-log(G_r(e^{i\omega})) \right | \]
\noindent  In such a situation

\begin{multline}
{J(\theta) \approx  \frac{1}{2\pi}\int_{-\pi}^{\pi}} \\
{ \left|   log(G_0(e^{i\omega}))-log(G_r(e^{i\omega}))       \right |^{2}\left | {B_r(e^{i\omega},\theta)} \right |^{2}\Psi_u(\omega)     d\omega}
\label{eq_4}
\end{multline}

\noindent or using a Stieltjes integral with $dQ_u(\omega)=\Psi_u(\omega)     d\omega$, for reasons that will become clear below
\begin{multline}
{J(\theta) \approx  \frac{1}{2\pi}\int_{-\pi}^{\pi}} \\
{ \left|   log(G_0(e^{i\omega}))-log(G_r(e^{i\omega}))       \right |^{2}\left | {B_r(e^{i\omega},\theta)} \right |^{2}dQ_u(\omega)}
\label{eq_generic_pb}
\end{multline}

\noindent This is what we call the initial criterion problem.

\section{Interpretation of the initial criterion}

\begin{figure}
	\centering
	\includegraphics[width=3in, keepaspectratio ]{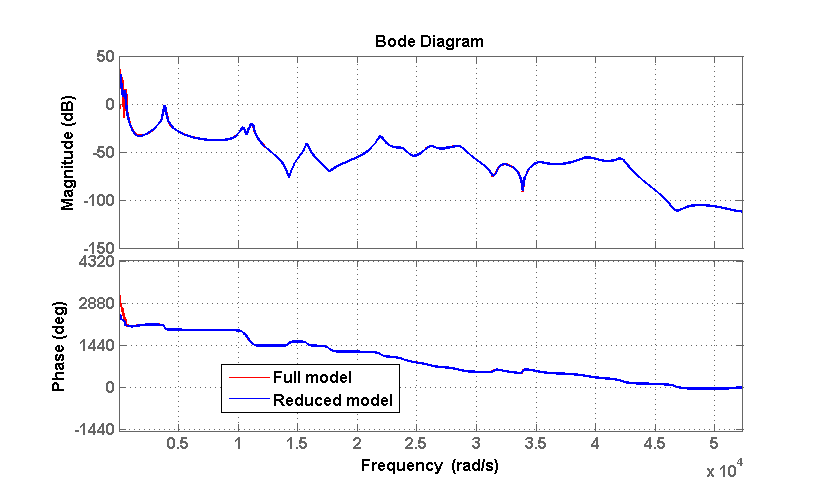}
	\caption{Classical Equation error reduction: results in a linear frequency scale}
	\label{fig1}
 \end{figure}

\begin{figure}
	\centering
	\includegraphics[width=3in, keepaspectratio ]{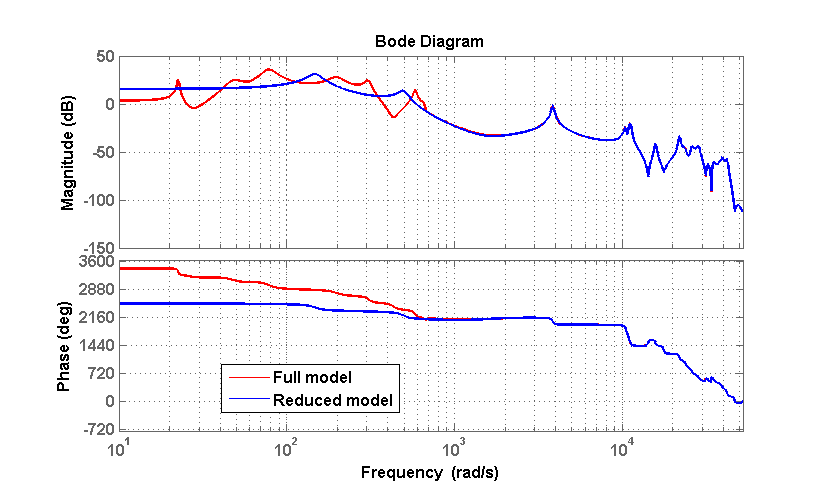}
	\caption{Classical Equation error reduction: results in a logarithmic frequency scale}
	\label{fig2}
 \end{figure}

\noindent For any frequency $\omega$, one can write
$G_0(e^{i\omega})=M_0(\omega)e^{i\varphi_0(\omega)}$
and
$G_r(e^{i\omega})=M_r(\omega)e^{i\varphi_r(\omega)}$,
where $M_0(\omega)$ and $M_r(\omega)$ denote the magnitudes,
and $\varphi_0(\omega)$ and $\varphi_r(\omega)$ denote the
phases of $G_0$ and $G_r$, respectively. As pointed out
in (\cite{r25}), one obtains
\begin{multline}
\left|
\log\left(G_0(e^{i\omega})\right)
-
\log\left(G_r(e^{i\omega})\right)
\right|^{2}
\\
=
\left|M_0(\omega)-M_r(\omega)\right|^{2}
+
\left|\varphi_0(\omega)-\varphi_r(\omega)\right|^{2}.
\label{eq_5}
\end{multline}
\noindent In \eqref{eq_5}, the discrepancies between the
systems in terms of magnitude and phase are clearly distinguished,
as can be visually observed in a Bode diagram.
\noindent Consequently, such a model reduction criterion
aims at minimizing the sum of the squared discrepancies
in magnitude and phase,
$\left|M_0(\omega)-M_r(\omega)\right|^{2}
+\left|\varphi_0(\omega)-\varphi_r(\omega)\right|^{2}$,
weighted by the term
$\left|B_r(e^{i\omega},\theta)\right|^2\Psi_u(\omega)$,
which depends on $\theta$.
One can infer that the term
$\left|B_r(e^{i\omega},\theta)\right|^2\Psi_u(\omega)$
will tend to be as small as possible, depending on the
flexibility of the reduced model, at frequencies where
$\left|M_0(\omega)-M_r(\omega)\right|^{2}
+\left|\varphi_0(\omega)-\varphi_r(\omega)\right|^{2}$
cannot be made small because of insufficient flexibility
in the reduced model. Conversely, if $G_0$ has a lightly damped zero at a
frequency $\omega_0$, minimizing
$\left|M_0(\omega_0)-M_r(\omega_0)\right|^{2}
+\left|\varphi_0(\omega_0)-\varphi_r(\omega_0)\right|^{2}$
calls for a lightly damped zero in $B_r$ at
$\omega=\omega_0$. This means that the discrepancy
$\left|M_0(\omega_0)-M_r(\omega_0)\right|^{2}
+\left|\varphi_0(\omega_0)-\varphi_r(\omega_0)\right|^{2}$
will be underweighted, and one can therefore expect
a poorer fit between the two models at $\omega_0$.
\noindent In any case, the effect of minimizing the
criterion can be properly interpreted only when the
Bode diagrams are considered with a frequency scale
consistent with the criterion, namely a linear frequency
scale (owing to the summation over $\omega$ and the
integration measure $d\omega$).
If one wishes to represent the system using Bode
diagrams with a logarithmic frequency scale, as is
usually the case, the criterion must be expressed
in terms of $\bar{\omega}=\log(\omega)$.
From \eqref{eq_generic_pb}, one immediately obtains
\begin{multline}
J(\theta) \approx \\
\frac{1}{\pi}
\int_{-\infty}^{\log(\pi)}
\left|
\log\left(G_0(e^{i e^{\bar{\omega}}})\right)
-
\log\left(G_r(e^{i e^{\bar{\omega}}})\right)
\right|^{2}
\\
\times
\left|
B_r(e^{i e^{\bar{\omega}}},\theta)
\right|^{2}
e^{\bar{\omega}}
\Psi_u(e^{\bar{\omega}})
\,d\bar{\omega}.
\label{eq_4}
\end{multline}
\noindent Owing to the weighting factor
$e^{\bar{\omega}}$, low frequencies are severely
underweighted in the integral. Consequently, poor
model matching can be expected at these frequencies,
making this method unsuitable for model reduction
in such frequency bands.
\noindent To illustrate these considerations, the
following example is presented. The full-order system
corresponds to the transfer function from the first
input to the second output of a MIMO benchmark
(SLICOT CD player) (\cite{r26}), whose order is $120$.
This full-order system, originally formulated in
continuous time, is discretized with a sampling
period of $6\times10^{-4}\,\mathrm{s}$ and excited
by a zero-mean white-noise sequence $\{u(t)\}$.
The classical equation error model reduction method
is applied (with the GOBF basis poles set to $0$),
using $n_a=48$.As shown in Fig.~\ref{fig1}, when the systems are
represented by Bode diagrams with a linear frequency
scale, the fit is very satisfactory. However, this
is no longer the case when a logarithmic frequency
scale is used, as illustrated in Fig.~\ref{fig2}.
In our opinion, this shortcoming explains why the
equation error method for model reduction, dating
back to (\cite{r27}), has seen such limited adoption.
The main issue is the mismatch between the frequency
scale on which the minimization is performed and
the frequency scale on which the system is intended
to be represented. To remedy this situation, the remainder of the paper
proposes to perform equation error model reduction
on an appropriate, tunable frequency scale, according
to the needs of the model user.

\section{The generalized equation error method}

Let us go back to the criterion \eqref{eq_crit_temp}. If the basis poles are no longer equal to $0$, one now obtains
\begin{multline}
{J(\theta)=\frac{1}{2\pi}\int_{-\pi}^{\pi}}\\
{\left|
G_r^{-1}(e^{i\omega},\theta)
\left(G_0(e^{i\omega})-G_r(e^{i\omega},\theta)\right)
\frac{B_r(e^{i\omega},\theta)}{A_p(e^{i\omega})}
\right|^{2}
\Psi_u(\omega)\,d\omega}
\label{eq_crit_with_GBOF2}
\end{multline}

\noindent Let us express this criterion in the Hambo frequency domain. At first, from Proposition 8 in (\cite{r2}),  the rational functions $\frac{A_r(z,\theta)}{A_p(z)}$ and $\frac{B_r(z,\theta)}{A_p(z)}$ have Hambo (system) transforms $\tilde{\mathcal{A}}_r(\lambda)$  and $\tilde{\mathcal{B}}_r(\lambda)$ which are polynomial matrices
\[ \tilde{\mathcal{A}}_r(\lambda)=\mathscr{A}_0+\mathscr{A}_1\lambda^{-1}+ \cdots +\mathscr{A}_n\lambda^{-n} \]
\[ \tilde{\mathcal{B}}_r(\lambda)=\mathscr{B}_0+\mathscr{B}_1\lambda^{-1}+ \cdots +\mathscr{B}_n\lambda^{-n} \]

\noindent thus the Hambo system transform of $G_r(z,\theta)$ denoted as $\tilde{G}_r(\lambda,\theta)$ is
\begin{equation}
 \tilde{G}_r(\lambda)=\tilde{\mathcal{A}}_r^{-1}(\lambda)\tilde{\mathcal{B}}_r(\lambda) 
 \label{eq_G_tilde}
 \end{equation}
\noindent  Set  $\tilde{G}_0 (\lambda)$ the Hambo system transform of $G_0(z)$. If $\tilde{G}_r(\lambda,\theta)$ is invertible, from Property \eqref{prop3}, one has 
\[ \tilde{G}^{-1}_r(\lambda)=\tilde{\mathcal{B}}_r^{-1}(\lambda)\tilde{\mathcal{A}}_r(\lambda) \]
\noindent and because of the commutativity Property \eqref{prop1}, one can use matrices logarithms as follows

\begin{equation}
 \tilde{G}_r^{-1}(\omega_\lambda) \left (\tilde{G}_0(\omega_\lambda)-\tilde{G}_r(\omega_\lambda)\right) \approx log (\tilde{G}_o(\omega_\lambda))-log (\tilde{G}_o(\omega_\lambda)) 
\label{eq_log_hambo}
\end{equation}
\noindent if
\[ \lVert  \tilde{G}_r^{-1}(\omega_\lambda) \left( \tilde{G}_0(\omega_\lambda)-\tilde{G}_r^{-1}(\omega_\lambda)\right) \rVert \ll 1\]

\noindent Now, setting 
\[ \breve{\Psi}_u(\omega_{\lambda})=\left(\Psi_u(\omega)\frac{1}{\beta^{'}(\omega)}\right)_{| \omega=\beta^{-1}(\omega_{\lambda})} \]
\noindent and
\begin{equation}
 d\breve{Q}(\omega_{\lambda})=\breve{\Psi}_u(\omega_{\lambda}))d\omega_{\lambda}= \Psi_u(\omega)d\omega=dQ(\omega)
 \label{eq_dQ_tilde}
 \end{equation}

\begin{figure}
    \centering
    \includegraphics[width=3.5in,keepaspectratio]{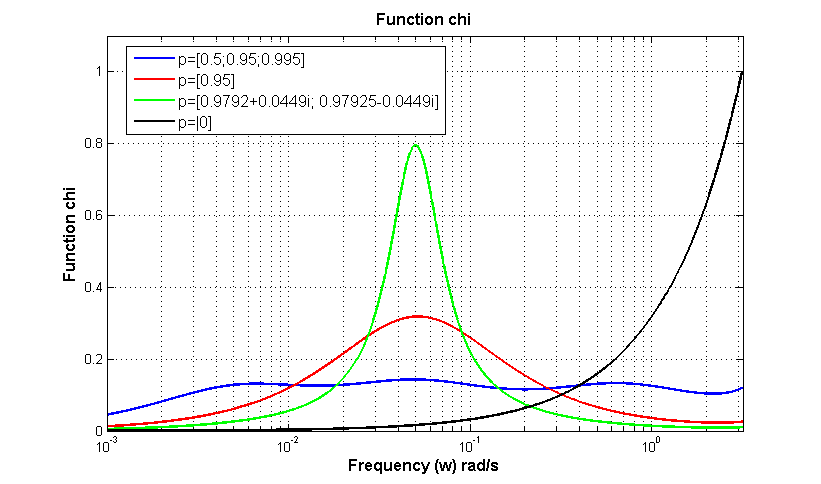}
    \caption{Examples of the function $\chi$ for different selections of basis poles.}
    \label{fig3}
\end{figure}

\begin{figure}[h]
    \centering
    \includegraphics[width=2.7in,keepaspectratio]{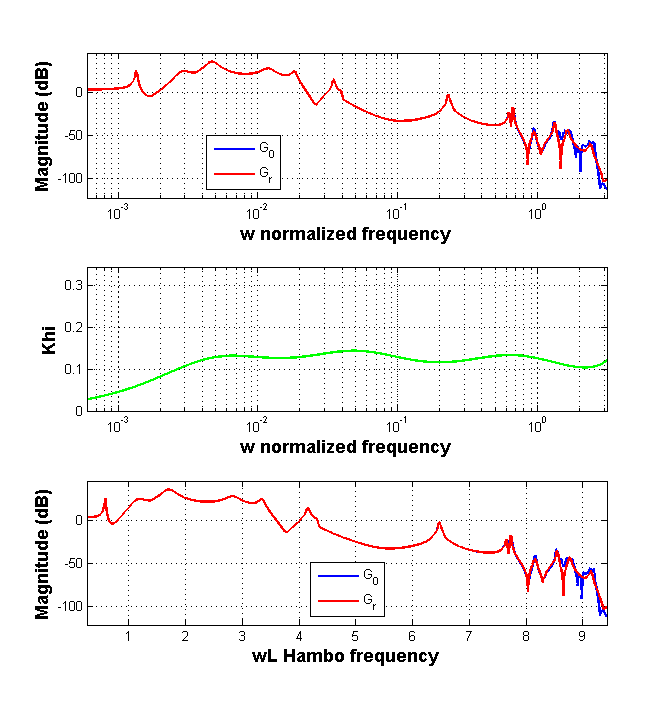}
    \caption{Model reduction results using a three-pole basis ($p_0=0.5$, $p_1=0.95$, $p_2=0.995$). Reduced-model order: $60$.}
    \label{fig4}
\end{figure}

\begin{figure}[h]
    \centering
    \includegraphics[width=2.7in,keepaspectratio]{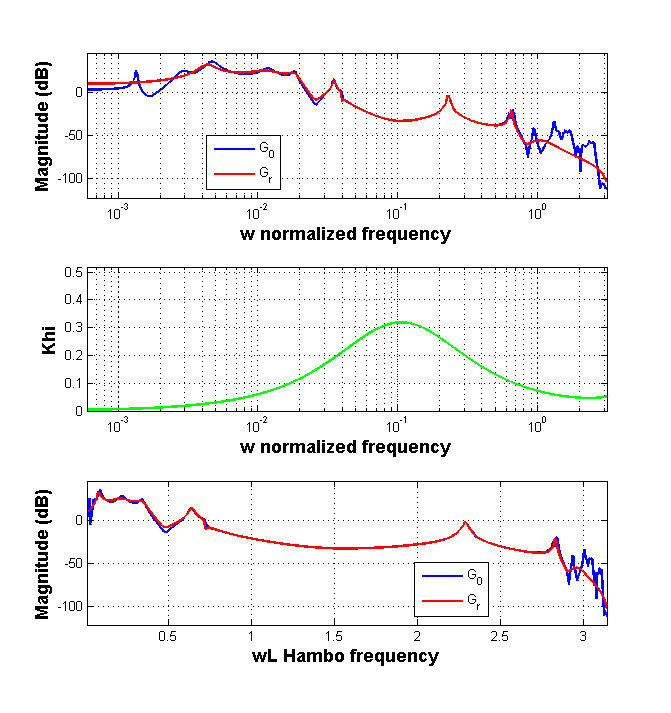}
    \caption{Model reduction results using a single-pole basis ($p_0=0.9$). Reduced-model order: $20$.}
    \label{fig5}
\end{figure}

\begin{figure}[h]
    \centering
    \includegraphics[width=2.7in,keepaspectratio]{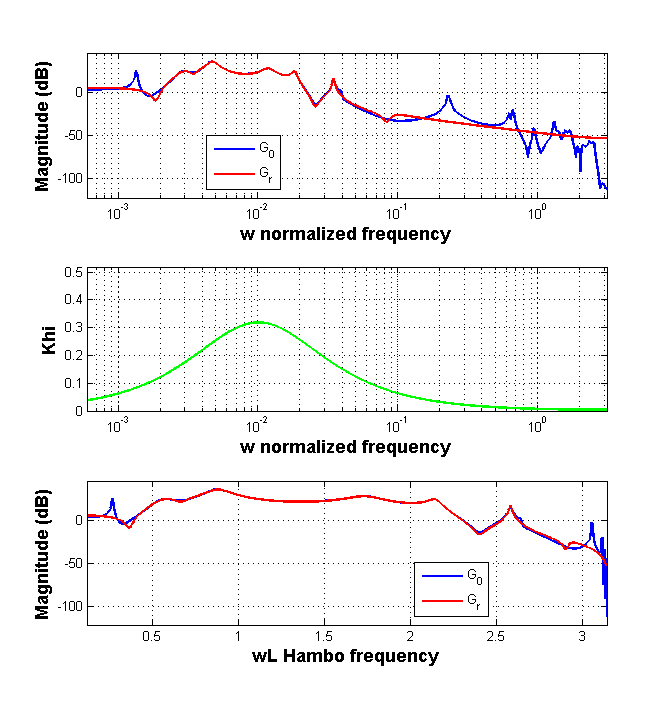}
    \caption{Model reduction results using a single-pole basis ($p_0=0.99$). Reduced-model order: $15$.}
    \label{fig6}
\end{figure}

\begin{figure}[h]
    \centering
    \includegraphics[width=2.7in,keepaspectratio]{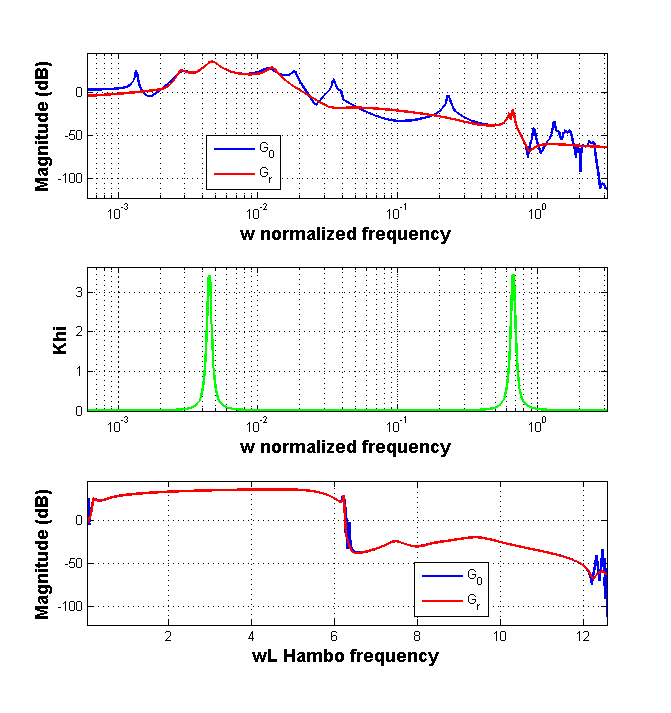}
    \caption{Model reduction results using a four-pole basis ($p_{0,1}=0.9997807 \pm 0.0044942i$ and $p_{3,1}=0.7602 \pm 0.6015i$). Reduced-model order: $20$.}
    \label{fig7}
\end{figure}

\noindent By combining \eqref{prop4}, \eqref{eq_G_tilde}, \eqref{eq_log_hambo}, and \eqref{eq_dQ_tilde}, one can now formulate the generic problem in the Hambo frequency domain, yielding

\begin{multline}
J \approx \frac{1}{2\pi n_p} \int_{-n_p \pi}^{n_p \pi}\\
\left\lVert
\left(
\log \tilde{G}_o(\omega_\lambda)
-\log(\tilde{G}_r(\omega_\lambda))
\right)
\tilde{\mathcal{B}}_r(\omega_\lambda)
\right\rVert_{2}^{2}
\,d\breve{Q}(\omega_{\lambda})
\label{eq_generic_pb_hambo}
\end{multline}
\noindent This equation should be compared with \eqref{eq_generic_pb}, which has exactly the same structure. Therefore, the use of GBOFs in the reduced-order model parameterization is equivalent to transposing the initial equation error criterion \eqref{eq_generic_pb} to the warped Hambo frequency scale. The full and reduced-order models are generally represented by Bode diagrams with a logarithmic frequency scale. Thus, the warping rate from the logarithmic frequency scale $\bar{\omega}$ to the Hambo frequency scale $\omega_\lambda$ provides a valuable heuristic tool for assessing the effect of basis-pole selection on the reduced-model fit. This warping rate has been studied in (\cite{r2}) (and also employed in (\cite{r33}) and is denoted by the function $\chi$. From the expression of the warping rate from the linear frequency scale $\omega$ to the Hambo frequency scale $\omega_\lambda$, recalled in \eqref{eq_beta_prime} (see Chapter 8 in (\cite{r1}) and (\cite{r24})), a normalized warping rate\footnote{In (\cite{r2}), (\cite{r33}), and (\cite{r22}), the term $n_p$ was omitted from the expression of the function $\chi$. However, this normalization term is necessary for the conservation principle to hold.} from $\bar{\omega}$ to $\omega_\lambda$ is defined, denoted by $\chi$ as in (\cite{r2}), such that
\begin{multline}
{\chi(e^{\bar{\omega}})=\frac{1}{n_p\pi}(e^{\bar{\omega}}) V_1^{T}(e^{ie^{\bar{\omega}}})V_1^{T}(e^{-ie^{\bar{\omega}}})=}\\
{\frac{1}{n_p\pi}(e^{\bar{\omega}})\sum_{j=0}^{n_p-1}\frac{1- \lvert p_j \rvert^{2}}{\lvert 1-\bar{p}_je^{ie^{\bar{\omega}}}\rvert^{2}}}
\label{eq_khi}
\end{multline}
\noindent The term $1/(n_p\pi)$ is introduced here solely for normalization purposes. Some important properties of the function $\chi$ are summarized below (for further details and proofs, see (\cite{r2})):
\begin{itemize}
\item A conservation principle holds:
\[
\int_{-\infty}^{\log(\pi)}
\chi(e^{\bar{\omega}})\,d\bar{\omega}=1.
\]
\item If a basis pole $p_j$ has a frequency sufficiently low compared with the Nyquist frequency, the function
$\bar{\omega} \mapsto \frac{1}{\pi}e^{\bar{\omega}}\frac{1- \lvert p_j \rvert^{2}}{\lvert 1-\bar{p}_je^{ie^{\bar{\omega}}}\rvert^{2}}$
has a local maximum located approximately at the frequency of the pole $p_j$.
\end{itemize}
\noindent Therefore, the shape of the function $\chi$ can be modified while preserving this conservation principle, with the basis poles acting as tuning parameters. In practical terms, the selection of these basis poles makes it possible to obtain a function $\chi$ that is
\begin{itemize}
\item spread over a very wide frequency range by selecting basis poles whose frequencies differ by several orders of magnitude (for example, $p_0=0.5$, $p_1=0.95$, $p_2=0.995$, as shown by the blue curve in Fig.~\ref{fig3});
\item concentrated within a frequency band of approximately one decade by selecting a single low-frequency basis pole, for example $p_0=0.95$ (red curve in Fig.~\ref{fig3});
\item concentrated within a very narrow frequency band by selecting two lightly damped poles, for example $p_{0,1}=0.9792 \pm 0.0449i$ (green curve in Fig.~\ref{fig3});
\item corresponding to the classical case, in which the reduced-order model is parameterized using $z^{-1}$, with a basis pole $p_0=0$. In this case, $\chi(e^{\bar{\omega}})=e^{\bar{\omega}}$, as shown by the black curve in Fig.~\ref{fig3}.
\end{itemize}
The purpose of the following simulations is to illustrate, on the one hand, the effect of basis-pole selection on the frequency bands in which the model fit is emphasized and, on the other hand, the usefulness of the function $\chi$ as a heuristic tool. The four examples presented below consider the same system as in Section 2 and illustrate the reduced-order models obtained for different selections of basis poles. In each figure, the magnitudes of $G_0$ and $G_r$ are displayed in the top plot as functions of frequency on a logarithmic scale. The middle plot represents the function $\chi$ on the same frequency scale, showing that the model fit is enhanced at frequencies where $\chi$ takes high values. The bottom plot of each figure compares $\|G_0(e^{i\omega})\|^2_{\omega=\beta^{-1}(\omega_{\lambda})}$ and $\|\tilde{G}_r(\omega_{\lambda})\|_2^2$ on the warped Hambo frequency scale ($\omega_{\lambda}=\beta(\omega)$), showing that the fit is uniformly distributed on this scale.
\noindent Fig.~\ref{fig4} presents the results obtained using a three-pole basis, for which the function $\chi$ is almost uniform over a wide frequency range. The reduced-model order is $60$.
\noindent The second example, presented in Fig.~\ref{fig5}, shows the reduced model obtained using a single-pole basis ($p_0=0.9$), with a reduced-model order of $20$.
\noindent The third example, shown in Fig.~\ref{fig6}, presents the reduced model obtained using a single-pole basis ($p_0=0.99$), with a reduced-model order of $15$.
\noindent Finally, the last example (Fig.~\ref{fig7}) considers a four-pole basis for which the function $\chi$ exhibits sharp peaks at $5\times10^{-3}\,\mathrm{rad/s}$ and $6.8\times10^{-1}\,\mathrm{rad/s}$ (normalized frequencies), corresponding to the basis poles $p_{0,1}=0.9997807 \pm 0.0044942i$ and $p_{3,1}=0.7602 \pm 0.6015i$. This illustrates that the model fit is concentrated within very narrow frequency bands around these frequencies. The reduced-model order is $20$.

\section{Towards the stochastic case}

\noindent In the case where the additive noise sequence $\{v(t)\}$ is nonzero, it is well known in the System Identification community that, in general, an additional undesirable bias appears. Specific identification schemes based on Pseudo-Linear Regression and established using GBOFs can be employed to be simultaneously insensitive to noise
effects and to allow the reduced-model fit to be managed through the selection of basis poles.
One can mention the generalized extended least-squares algorithm, dedicated to ARMAX systems, called H-ELS
in (\cite{r22}). A whiteness test, also established using GBOFs and consistent with H-ELS, has been proposed
in (\cite{r33}) to validate a reduced model over a specific frequency band. A complete framework for data-driven
model reduction using GBOF-based algorithms will be presented in a future publication.


\begin{thebibliography}{99}       

\bibitem [Antoulas (2005) ]{r29} A.C. Antoulas \textit{Approximation of large-scale dynamical systems}, SIAM, 2005.                                            
\bibitem [Benner \textit{et al.} (2015) ]{r28} P.Benner, S.Gugercin, K.Willcox, "A survey of projection-based model reduction methods for parametric dynamical systems", SIAM Review, vol.57(4), pp.483-531, 2015.

\bibitem [Benner \textit{et al.} (2023)]{r32} P. Benner, P. Goyal, B. Kramer, B. Peherstorfer, K. Willcox, "Operator inference for nonintrusive model reduction of systems governed by parametric PDEs", Computer methods in applied mechanics and engineering, vol.403(B), 2023.

\bibitem [ Chahlaoui,Van Dooren (2002)]{r26}  Y. Chahlaoui, P. Van Dooren, "A collection of benchmark examples for model reduction of linear time invariant dynamical systems", Slicot working note 2002-2, on the internet.


\bibitem [Heuberger \it{et al.} (2005) ] {r1} P.S.C. Heuberger, P.M.J Van den Hof, B. Wahlberg, \textit{Modelling and identification with rational orthogonal basis functions}, Springer Verlag, 2005.       

\bibitem [Mayo, Antoulas (2007) ]{r30}  A.J. Mayo, A.C. Antoulas, "A framework for the solution of the generalized realization problem", vol.425(2-3), pp.634-662. Linear Algebra and its applications, 2007.

\bibitem  [Mullis, Roberts (1976)] {r27} C. T. Mullis, R. A. Roberts, "The use of second-order information in the approximation of discrete-time linear systems", IEEE Transactions on acoustics, speech and signal processing, vol. 24(3), 1976.

\bibitem[Shipp \textit{et al.} (1996)]{r24} F. Shipp, L. Gianone, J. Bokor, Z. Szabo, "Identification in generalized orthogonal basis - a frequency domain approach", preprints of the 13th IFAC World congress, vol. 1, pp. 387-392,1996.

\bibitem [Tu \textit{et al.} (2014) ]{r31} J.H. Tu, C.W. Rowley, "On dynamic decomposition: Theory and applications", Journal of  computational dynamics, vol.1(2), 2014.

\bibitem [Vau, Bourlès (2021a)] {r2} B. Vau, H. Bourlès, "Closed-loop output error identification algorithms with predictors based on generalized orthonormal transfer functions: Convergence conditions and bias distribution", Automatica (Elsevier), vol. 125, March 2021.


\bibitem [Vau, Bourlès (2022) ]{r33} B. Vau, H. Bourl\`es, A robust whiteness test for the identification of discrete-time linear models: Use of orthonormal transfer functions, Automatica, vol.139, 2022.

\bibitem [Vau, Bourlès (2026) ]{r22} B. Vau, H. Bourl\`es, "On the bias distribution of pseudo-linear regression algorithms for recursive identification", International Journal of Adaptive Control and Signal Processing, 2026.

\bibitem [Wahlberg, Ljung (1986)] {r25} B. Wahlberg, L. Ljung, "Design variables for bias distribution in transfer function estimation", IEEE Trans. Aut. Control, vol. 31(2), pp. 134-144, 1986.





\end{thebibliography}
\end{document}